\documentclass[%
 reprint,
 amsmath,amssymb,
 aps,
]{revtex4-2}

\usepackage{graphicx}
\usepackage{dcolumn}
\usepackage{bm}
\usepackage{subfigure}
\usepackage[dvipsnames]{xcolor}
\usepackage{amssymb,mathtools}
\usepackage{bigints}

\begin{document}

\preprint{APS/123-QED}

\title{Characteristics and Implementation of B$\boldsymbol{^{\alpha}}$ Gates}

\author{M. Karthick Selvan}
\email{karthick.selvan@yahoo.com}%

\author{S. Balakrishnan}%
 \email{physicsbalki@gmail.com}
\affiliation{Department of Physics, School of Advanced Sciences, Vellore Institute of Technology, Vellore - 632014, Tamilnadu, India.}%



\begin{abstract}
 In this brief report, we discuss the characteristics of B$^{\alpha}$ gates. We provide the conditions for the two-qubit gates generated by two applications of a B$^\alpha$ gate. We propose an experimental scheme to implement B$^{\alpha}$ gates in ion-trap system. In this scheme, we assume that only a single vibrational mode contributes to spin-spin coupling. This scheme is an extension of a recently proposed scheme to realize XY-type interaction in ion-trap system. With the successful implementation of this scheme, B$^\alpha$ gates can be used for doing quantum computation in ion-trap quantum computers. 
\end{abstract}

\maketitle

Entangling two-qubit gates play a major role in quantum computation and quantum information processing. Fixed basis gates set model of quantum computation uses a specific entangling two-qubit gate and a set of single-qubit gates to perform quantum computation. In this model, the unitary operations that will be performed on qubits are fixed and hence the operating parameters of quantum devices, performing those unitary operations, can also be fixed. A suitable entangling basis gate is the one that can be implemented with high fidelity and implements an arbitrary unitary operation in the least number of applications. Understanding the geometry and nonlocal characteristics of two-qubit gates are necessary to choose a suitable entangling basis gate. 

A two-qubit gate, $U \in U(4)$, can be decomposed as follows~\cite{Zhang2003}. 
\begin{equation}\label{eq1}
U = e^{i \phi} (k_1 \otimes k_2) U_d (c_1, c_2, c_3) (k_3 \otimes k_4).
\end{equation}
where $e^{i \phi}$ is the global phase, $k_{1(3)} \otimes k_{2(4)} \in SU(2) \otimes SU(2)$ are local gates, $U_d(c_1, c_2, c_3) = \exp \left[ i \left( c_1 (\sigma_x \otimes \sigma_x) + c_2 (\sigma_y \otimes \sigma_y) + c_3 (\sigma_z \otimes \sigma_z) \right)/2 \right] \in SU(4)$ is the nonlocal part of two-qubit $U$ and $(c_1, c_2, c_3)$ are Cartan co-ordinates. 

\begin{figure}[h]
\includegraphics[width=0.5\textwidth]{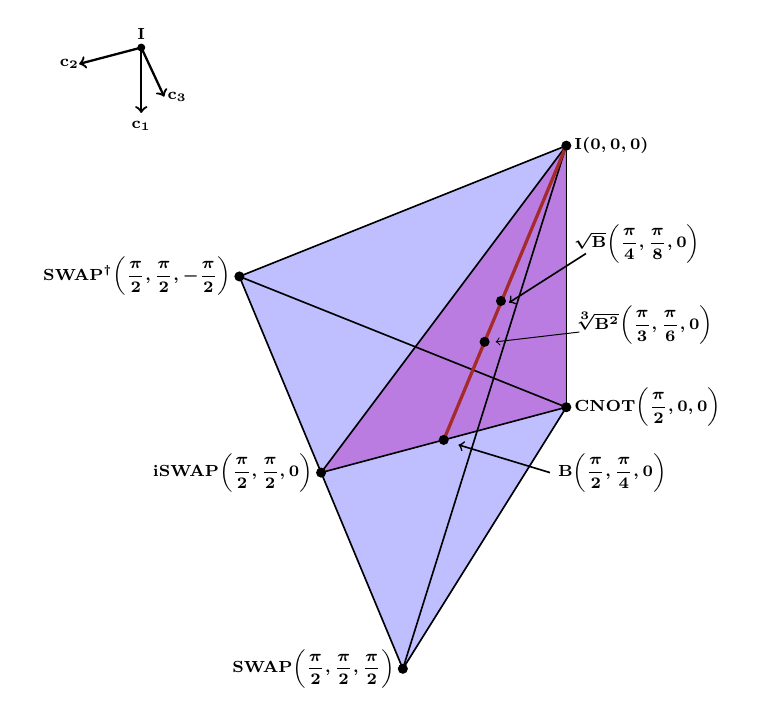}
\caption{Geometry of local equivalence classes of two-qubit gates}
\label{fig1}
\end{figure}

Two-qubit gates differing by local operations form a local equivalence class and the local equivalence classes of two-qubit gates are geometrically represented by the points of a tetrahedron shown in FIG.~\ref{fig1}. Each point of this tetrahedron is the Cartan co-ordinates of a local equivalence class of two-qubit gates and they satisfy the condition: $\pi/2 \geq c_1 \geq c_2 \geq \vert c_3 \vert \geq 0$. Local invariants of two-qubit gates which can be used to distinguish the two-qubit gates belonging to different local equivalence classes can be written in terms of Cartan co-ordinates as follows~\cite{Watts2013}. 
\begin{equation*}
G_1 = \dfrac{1}{4} \left[\sum_{p=1}^3 \cos(2c_p)  + \prod_{q=1}^3 \cos(2c_q) + i \prod_{r=1}^3 sin(2c_r) \right], 
\end{equation*}
\begin{equation}\label{eq2}
G_2 = \sum_{p=1}^3 \cos(2c_p).
\end{equation}
Entangling power~\cite{Zanardi2000}, the average entanglement generated over all product states by a two-qubit gate can be expressed in terms of its Cartan co-ordinates as follows~\cite{Rezakhani2004}. 
\begin{equation}\label{eq3}
e_p (c_1, c_2, c_3) = \dfrac{1}{18} \left[ 3 - \dfrac{1}{2} \sum_{i,j=1, i \neq j}^3 \cos(2c_i) \cos(2c_j) \right].
\end{equation}
Two-qubit gates having maximum entangling power are called special perfect entanglers (SPEs). SPEs are represented by the line connecting CNOT$(\pi/2, 0, 0)$ and iSWAP$(\pi/2, \pi/2, 0)$ in FIG.~\ref{fig1}. 
The family of B$^{\alpha}$ gates is represented by the line connecting $\text{I}(0,0,0)$ and $\text{B}(\pi/2, \pi/4, 0)$ in  FIG.~\ref{fig1}. Cartan co-ordinates of B$^{\alpha}$ gates are $\left( \alpha \pi/2, \alpha \pi/4, 0 \right)$ with $\alpha \in [0,1]$. Their local invariants are $G_1 = \left[ \left( 1 + \cos \left( \alpha \pi \right) \right) \left( 1 +  \cos \left( \alpha \pi/2 \right) \right) \right]/4$ and $G_2 = 1 + \cos \left( \alpha \pi \right) + \cos \left( \alpha \pi/2 \right)$. The entangling power of B$^{\alpha}$ gates is given by 
\begin{equation}\label{eq4}
e_p \left( \text{B}^{\alpha} \right) = \dfrac{1}{18} \left[3 - \cos \left( \dfrac{\alpha \pi}{2} \right) \left[ \cos(\alpha \pi) + 1 \right]  - \cos(\alpha \pi)  \right].
\end{equation}
Two noteworthy gates belonging to this family are B gate and $\sqrt{\text{B}}$ gate. B gate is an SPE and it is the only two-qubit gate that can generate any two-qubit gate in two applications~\cite{Zhang2004}. Hence, it can be shown that for every two-qubit gate $U$, there exists a gate B$'$ in the local equivalence class of B gate such that $U^\dagger \text{B}' U$ is also in the local equivalence class of B gate. The average number of applications required to generate Haar random two-qubit gates is smaller for $\sqrt{\text{B}}$ gate as compared to CNOT gate~\cite{McKinney2023}. $\sqrt{\text{B}}$ gate has been proposed as an alternative to CNOT gate as an entangling basis gate~\cite{Selvan2023,Selvan2024}. Another interesting gate among B$^{\alpha}$ family is $\sqrt[3]{\text{B}^2}$ gate. This gate is represented by the centroid of $c_3 = 0$ plane in FIG.~\ref{fig1}. The set of gates that can be generated by two applications of a B$^\alpha$ gate satisfy either 

\begin{equation*}
c_1 - c_2 \geq 0
\end{equation*}
\begin{equation*}
c_2 - c_3 \geq 0
\end{equation*}
\begin{equation*}
c_3 \geq 0
\end{equation*}
\begin{equation*}
\alpha \pi \geq c_1 
\end{equation*}
\begin{equation}\label{eq5}
\dfrac{3 \alpha \pi}{2} \geq c_1 + c_2 + c_3
\end{equation}

or 

\begin{equation*}
(\pi - c_1) - c_2 \geq 0
\end{equation*}
\begin{equation*}
c_2 - c_3 \geq 0
\end{equation*}
\begin{equation*}
c_3 \geq 0
\end{equation*}
\begin{equation*}
\alpha \pi \geq (\pi - c_1) 
\end{equation*}
\begin{equation}\label{eq6}
\dfrac{3 \alpha \pi}{2} \geq (\pi - c_1) + c_2 + c_3
\end{equation}

\begin{figure}[h]
\begin{tabular}{c}
\includegraphics[width=0.5\textwidth]{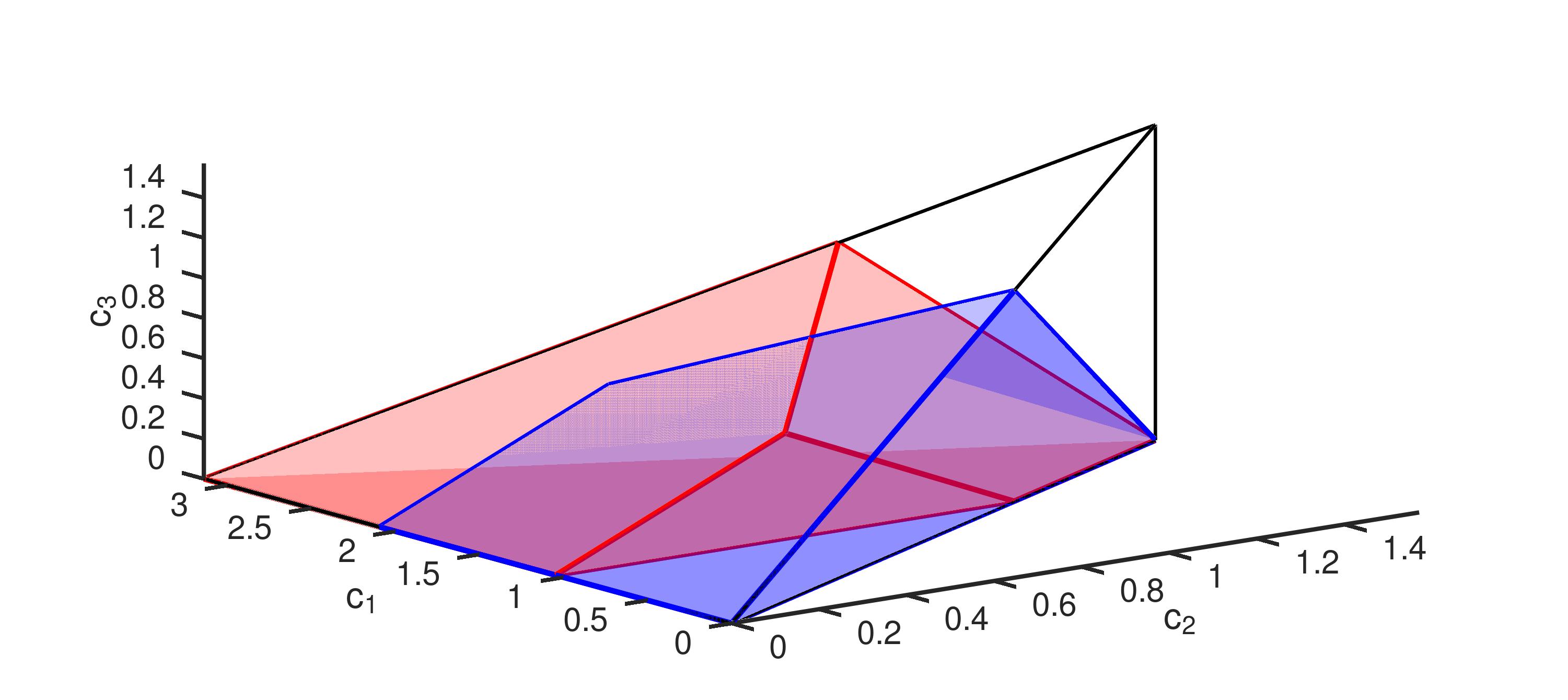} 
\end{tabular}
\caption{Weyl chamber coverage of two applications of $\sqrt[3]{\text{B}^2}$ gate in positive Cartan coordinate system.}
\label{fig2}
\end{figure}

These two sets of inequality relations are obtained using the analytical method described in Ref.~\cite{Peterson2020}. The co-ordinates in the two sets of inequalities, $(c_1, c_2, c_3)$, describe positive Cartan co-ordinate system. The first three inequalities, in both sets, describe the Weyl chamber and the last two conditions describe the region within Weyl chamber covered by the gates that can be generated by two applications of the B$^\alpha$ gate. For $\alpha = 2/3$, the region of Weyl chamber covered by the gates satifying the above conditions are shown in FIG.~\ref{fig2}. In FIG.~\ref{fig2}, the region blue (red) in color represents the gates that satisfy the first (second) set of inequality relations. It can be verified that three applications of $\sqrt[3]{\text{B}^2}$ gate generate all two-qubit gates.  

Now we discuss about the implementation of B$^\alpha$ gates. A scheme to implement $\sqrt{\text{B}}$ gate in ion-trap system has been mentioned~\cite{Selvan2023}. It involves exciting two ions with a pair of oppositely detuned laser fields. In the weak coupling regime~\cite{Sorensen1999,Sorensen2000}, one can have pure $\sigma_x \otimes \sigma_x$ or $\sigma_y \otimes \sigma_y$ interactions. B$^{\alpha}$ gates can be implemented in ion-trap system by allowing $\sigma_x \otimes \sigma_x$ and $\sigma_y \otimes \sigma_y$ interactions sequentially for specific durations. Recently, in ion-trap system, simultaneous realization of $\sigma_x \otimes \sigma_x$ and $\sigma_y \otimes \sigma_y$ interactions using two pairs of laser fields in the weak coupling regime is proposed~\cite{Kotibhaskar2023}. This scheme can be adapted to apply two-qubit XY gates between pair of ions and B$^\alpha$ gates can be implemented by choosing appropriate Rabi frequencies. But, in the weak coupling regime, implementation of gates would be slower. Hence, in this paper, we extend the simultaneous scheme~\cite{Kotibhaskar2023} to a general case where only a single vibrational mode is assumed to contribute to spin-spin coupling~\cite{Sorensen2000,Kim2009} and we show that this scheme is more suitable to implement B$^{\alpha}$ gates.  

We consider two internal states, ground state $(\vert g \rangle)$ and excited state $(\vert e \rangle)$ with energy separation $\hbar \omega_{eg}$, of ions as qubits. Qubits are coupled via their vibrational motion. We consider two pairs of laser fields, one pair with frequencies $\omega_{eg} \pm \mu_x$ and another pair with frequencies $\omega_{eg} \pm \mu_y$. The detunings $\mu_x$ and $\mu_y$ are chosen to be close to a particular vibrational mode frequency $\omega$ and we assume that contribution of other vibrational modes are negligible. The phases of laser fields are chosen such that one pair applies $\sigma_x$ force and another applies $\sigma_y$ force. Under Lamb-Dicke and rotating wave approximations, the interaction of ions with laser fields is described by the following Hamiltonian. 
\begin{equation}
H \left( t \right) = H_x \left( t \right) + H_y \left( t \right),
\end{equation}
with
\begin{equation}
H_x \left( t \right) = \dfrac{1}{2} \sum_m \eta_m \Omega_m^x \sigma_x^m \left[ a^{\dagger} e^{-i \left( \delta_x t + \psi_x \right)} + a e^{i \left( \delta_x t + \psi_x \right)}\right],
\end{equation}
and 
\begin{equation}
H_y \left( t \right) = \dfrac{1}{2} \sum_n \eta_n \Omega_n^y \sigma_y^n \left[ a^{\dagger} e^{-i \left( \delta_y t + \psi_y \right)} + a e^{i \left( \delta_y t + \psi_y \right)}\right],
\end{equation}
where $\eta_{m(n)}$ is Lamb-Dicke parameter, $\Omega_{m(n)}^{x(y)}$ is Rabi frequency, $\psi_{x(y)}$ is motional phase and $\delta_{x(y)} = \mu_{x(y)} - \omega$. 

Time evolution of the system can be described by the operator obtained from Magnus expansion terminated with two terms. 
\begin{equation}
U \left( \tau \right) = \exp \left[ - i \int_0^\tau H \left( t \right) dt - \dfrac{1}{2} \int_0^\tau dt_1 \int_0^{t_1} [ H \left( t_1 \right) , H \left( t_2 \right)] dt_2 \right].
\end{equation}
After substituting the expression of Hamiltonian and doing the integration, the time evolution operator can be expressed as follows.  
\begin{widetext}
\begin{equation*}
U \left( \tau \right) = \exp \bigg[ \sum_m \zeta_m^x \left( \tau \right) \sigma_x^m  + \sum_n \zeta_n^y \left( \tau \right) \sigma_y^n - i \sum_{m , n} \chi^x_{m,n} \left(\tau \right) \sigma_x^m \sigma_x^n - i \sum_{m , n} \chi^y_{m,n} \left( \tau \right) \sigma_y^m \sigma_y^n 
\end{equation*}
\begin{equation}
- i \sum_{m,n} \left[ \Lambda_{m,n} \left( \tau \right)  \sigma_y^m \sigma_x^n + \Lambda'_{m,n} \left( \tau \right) \sigma_x^m \sigma_y^n \right]+ i \sum_m \xi_m \left( \tau \right) \sigma_z^m \bigg],
\end{equation}
where 
\begin{equation}
\zeta_{m (n)}^{x (y)} ( \tau) = \alpha_{m (n)}^{x(y)} \left( \tau \right) a^{\dagger} - \alpha^{x(y)*}_{m (n)} \left( \tau \right) a,~~~\text{with}~~~\alpha_{m(n)}^{x(y)} (\tau) = \dfrac{\eta_{m(n)} \Omega^{x(y)}_{m(n)}e^{-i\psi_{x(y)}}}{2 \delta_{x(y)}} \left[ e^{-i \delta_{x(y)} \tau} - 1 \right],
\end{equation}
\begin{equation}
\chi^{x(y)}_{m,n} (\tau) = \dfrac{\eta_m \eta_n \Omega_m^{x(y)} \Omega_n^{x(y)}}{4 \delta^2_{x(y)}} \left[ \delta_{x(y)} \tau - \sin \left( \delta_{x(y)} \tau \right)  \right],
\end{equation}
\begin{equation}
\Lambda_{m,n} \left( \tau \right) = \dfrac{\eta_m \eta_n \Omega_m^y \Omega_n^x}{4} \left[ \dfrac{\sin \left[ \left(\delta_y - \delta_x \right) \tau + \psi_y - \psi_x \right] - \sin \left[\psi_y - \psi_x \right]}{\left(\delta_y - \delta_x \right) \delta_x} - \dfrac{\sin \left[ \delta_y \tau + \psi_y - \psi_x \right] - \sin \left[ \psi_y - \psi_x \right]}{\delta_x \delta_y} \right],
\end{equation}
\begin{equation}
\Lambda'_{m,n} \left( \tau \right) = \dfrac{\eta_m \eta_n \Omega_m^x \Omega_n^y}{4} \left[ \dfrac{\sin \left[ \left(\delta_x - \delta_y \right) \tau + \psi_x - \psi_y \right] - \sin \left[\psi_x - \psi_y \right]}{\left(\delta_x - \delta_y \right) \delta_y} - \dfrac{\sin \left[ \delta_x \tau + \psi_x - \psi_y \right] - \sin \left[ \psi_x - \psi_y \right]}{\delta_x \delta_y} \right],
\end{equation}
and
\begin{equation}
\xi_m \left( \tau \right) = \beta^1_m \left( \tau \right) {a^{\dagger}}^2 + \beta^2_m \left( \tau \right) a^{\dagger} a + \text{h.c.},
\end{equation}
with 
\begin{equation}
\beta^1_m \left( \tau \right) = \dfrac{\eta_m^2 \Omega_m^x \Omega_m^y}{4} \left[ \dfrac{1 - e^{-i (\delta_x + \delta_y ) \tau}}{( \delta_x + \delta_y)} \left( \dfrac{1}{\delta_x} - \dfrac{1}{\delta_y} \right) - \dfrac{e^{-i \delta_x \tau} - e^{-i \delta_y \tau}}{\delta_x \delta_y}\right] e^{-i (\psi_x + \psi_y)},
\end{equation}
and 
\begin{equation}
\beta^2_m \left( \tau \right) = \dfrac{\eta_m^2 \Omega_m^x \Omega_m^y}{4} \left[ e^{-i(\psi_y - \psi_x )} \left( \dfrac{e^{-i(\delta_y - \delta_x) \tau} - 1}{\delta_x (\delta_y - \delta_x)} - \dfrac{e^{-i \delta_y \tau} - 1}{\delta_x \delta_y} \right) - 
e^{-i(\psi_x - \psi_y )} \left( \dfrac{e^{-i(\delta_x - \delta_y) \tau} - 1}{\delta_y (\delta_x - \delta_y)} - \dfrac{e^{-i \delta_x \tau} - 1}{\delta_x \delta_y} \right) \right].
\end{equation}
\end{widetext}
In time evolution operator, the first two terms describe spin-phonon couplings, the second two terms describe pure spin-spin couplings and the last two terms describe spin-spin and spin-phonon couplings due to cross terms $\left[ H_x(t_1), H_y(t_2) \right]$ and $\left[ H_y(t_1), H_x(t_2) \right]$. Choosing the detunings $\mu_{x(y)}$ and gate time $\tau$ such that $\delta_x \tau = 4 N \pi$ and $\delta_y \tau = 2 N \pi$ (for integer $N$), the angles $\zeta_{m(n)}^{x(y)}$'s, $\Lambda_{mn}$'s, $\Lambda'_{mn}$'s, and $\xi_m$'s can be made zero at the end of gate time $\tau$. The spin-spin coupling angles $\chi_{m,n}^{x(y)}(\tau)$ between pair of ions become 
\begin{equation}
\chi_{m,n}^x(\tau) = \dfrac{2 \pi N \eta_m \eta_n \Omega_m^x \Omega_n^x }{\delta_x^2},
\end{equation}
\begin{equation}
\chi_{m,n}^y(\tau) = \dfrac{\pi N \eta_m \eta_n \Omega_m^y \Omega_n^y }{\delta_y^2}.
\end{equation}
The natural interaction implementing B$^{\alpha}$ gates is given by 
\begin{equation}
H_{\text{B}^{\alpha}} = \dfrac{\alpha \pi}{2} \left( \sigma_x \otimes \sigma_x \right) + \dfrac{\alpha \pi}{4} \left( \sigma_y \otimes \sigma_y \right).
\end{equation}
This interaction can be realized by choosing $\Omega_m^x \Omega_n^x = 4 \Omega_m^y \Omega_n^y$. Comparing the coupling angles in $\exp\left( -i H_{\text{B}^\alpha}/2 \right)$ with those given in Eqs. 17 and 18, we get
\begin{equation}
\dfrac{\eta_m \eta_n \Omega_m^x \Omega_n^x}{\delta_x^2} = \dfrac{\alpha}{8N}.
\end{equation}
Thus by changing the Rabi frequencies $\Omega_m^{x(y)}$ and $\Omega_n^{x(y)}$ such that $(\Omega_m^x \Omega_n^x)/(\Omega_m^y \Omega_n^y) = 4 $, the desired B$^\alpha$ gate can be implemented. Using this scheme, B$^\alpha$ gates can be used as native gates in ion-trap system. Rabi frequencies can be modified by changing the power of laser fields. Implementation of $\sqrt{\text{B}}$ gate requires lesser power than that required to implement B gate. Hence, $\sqrt{\text{B}}$ gate can be implemented with better fidelity and a fixed basis gates set can be formed using $\sqrt{\text{B}}$ gate as an entangling basis gate for doing quantum computation in ion-trap quantum computers.


\end{document}